\overfullrule=0pt
\input harvmac
\def\e{{\epsilon}}
\def\s{{\sigma}}
\def\N{{\nabla}}
\def\O{{\Omega}}

\def\half{{1\over 2}}
\def\p{{\partial}}

\def\t{{\theta}}
\def\ta{{\theta^{\bar a}}}
\def\tb{{\theta^{\bar b}}}

\Title{\vbox{\hbox{IFT-P.013/99 }}}
{\vbox{
\centerline{\bf Quantization of the Superstring with}
\centerline{\bf  Manifest U(5) Super-Poincar\'e Invariance}}}

\bigskip\centerline{Nathan Berkovits\foot{e-mail: nberkovi@ift.unesp.br}}
\bigskip
\centerline{\it Instituto de F\'\i sica Te\'orica, Universidade Estadual
Paulista}
\centerline{\it Rua Pamplona 145, 01405-900, S\~ao Paulo, SP, Brasil}

\vskip .3in
The superstring is quantized in a manner 
which manifestly preserves a U(5) subgroup of the (Wick-rotated) 
ten-dimensional super-Poincar\'e invariance. This description 
of the superstring contains critical N=2 worldsheet superconformal 
invariance and is a natural covariantization of the U(4)-invariant 
light-cone Green-Schwarz description.

\Date {February 1999}

\newsec{Introduction}

Although the uncompactified superstring is invariant under
SO(9,1) super-Poincar\'e transformations, there is no quantizable formalism
for the superstring where all of these invariances can be made
manifest.
In the conventional Ramond-Neveu-Schwarz (RNS) formalism, the bosonic 
SO(9,1) Poincar\'e invariances can be made manifest for computations
involving external Neveu-Schwarz states. For computations
involving external Ramond states, the necessity of bosonizing
the ten $\psi$ fields breaks the SO(9,1) invariance to
U(5) (after Wick-rotating from SO(9,1) to SO(10)).
Furthermore, none of the spacetime-supersymmetries are manifest
in the RNS formalism. 

In the light-cone Green-Schwarz formalism for the superstring, only a U(4)
subgroup of the super-Poincar\'e invariances can be made manifest.\foot
{Although physical states in light-cone gauge
can be described in an SO(8)-invariant
manner, amplitude computations require that four of
the eight $\theta$'s carry spin-0 and the other four carry spin-1,
breaking SO(8) down to U(4).\ref\lcgs{M.B. Green and J.H. Schwarz,
Nucl. Phys. B243 (1984) 475\semi
S. Mandelstam, Prog. Theor. Phys. Suppl. 86 (1986) 163\semi
A. Restuccia and J.G. Taylor, Phys. Rep. 174 (1989) 283.}}
Although this U(4) subgroup includes some spacetime supersymmetries, 
the light-cone Green-Schwarz formalism is completely gauge-fixed 
(i.e. has no worldsheet conformal
or superconformal invariances) which leads to technical
difficulties when computing
scattering amplitudes.
There also exists a classical covariant version of the Green-Schwarz
formalism which has manifest SO(9,1) super-Poincar\'e invariance, but it
has not yet been quantized in a manner suitable for defining physical
states or computing scattering amplitudes.

Over the last five years, a new formalism for the superstring has
been developed which is a hybrid between the RNS and Green-Schwarz 
formalisms. It contains some manifest spacetime-supersymmetries, 
and is related to the conventional RNS formalism by a redefinition
of the worldsheet variables. The formalism contains critical N=2 worldsheet 
superconformal invariance, and amplitudes can be computed using 
standard N=2 string methods or using N=4 topological string methods.
In previous papers, it was shown how to perform a field-redefinition from
RNS variables to Green-Schwarz-like variables 
which preserves manifest SO(3,1)
super-Poincar\'e invariance\ref\four
{N. Berkovits, ``Covariant Quantization Of
The Green-Schwarz Superstring in A Calabi-Yau Background,''
Nucl. Phys. {\bf B431} (1994) 258\semi
N. Berkovits, ``A New Description Of The Superstring,''
Jorge Swieca Summer School 1995, p. 490, hep-th/9604123.}
or manifest
$SO(5,1)$ super-Poincar\'e invariance
\ref\Topo{N. Berkovits and C. Vafa,
``$N=4$ Topological Strings'', Nucl. Phys. B433 (1995) 123, hep-th/9407190
\semi N. Berkovits, C. Vafa and E. Witten, ``Conformal Field
Theory of AdS Background with Ramond-Ramond Flux'', hep-th/9902098.}.
In both these cases, four of the sixteen spacetime supersymmetries
are manifest. 

In this paper, it will be shown how to perform a field-redefinition
of the RNS worldsheet variables which manifestly preserves 
six of the sixteen
spacetime-supersymmetries. After Wick-rotating to SO(10), this 
formalism contains manifest U(5) super-Poincar\'e invariance. 
(Before Wick-rotating, there is a corresponding non-compact
subgroup of SO(9,1) containing 25 bosonic generators and 6 fermionic 
generators which is manifest.) 
So this formalism
contains the same manifest bosonic symmetries as the RNS formalism 
in the presence
of Ramond states, 
but it also contains six manifest spacetime supersymmetries.
Furthermore, one can define 
the physical state conditions in a manner which is 
manifestly invariant under these symmetries.

It is interesting to note that the U(5)
transformations (or their Minkowski counterpart) preserve up to
rescaling a pure spinor, i.e. 
a bosonic complex
spinor $u^A$ satisfying $u^A \sigma^m_{AB} u^B=0$ for $m=0$ to 9.
Pure spinors have been suggested by other authors to play
a useful role in describing $d=10$ super-Yang-Mills and supergravity
in harmonic superspace\ref\pure{B.E.W. Nilsson,
``Pure Spinors as Auxiliary Fields in the Ten-Dimensional
Supersymmetric Yang-Mills Theory'', Class. Quant. Grav. 3 (1986) L41\semi
P.S. Howe, ``Pure Spinor Lines in
Superspace and Ten-Dimensional Supersymmetric Theories'', Phys. Lett.
B258 (1991) 141\semi P. Howe, ``Pure Spinors, Function Superspaces and
Supergravity Theories in Ten and Eleven Dimensions'', Phys. Lett. B273
(1991) 90.}.

\newsec{Review of Spacetime Supersymmetry in the RNS Formalism}

As shown in 
\ref\emb{N. Berkovits, 
``The Ten-Dimensional Green-Schwarz Superstring is a
Twisted Neveu-Schwarz-Ramond String'', Nucl. Phys. B420 (1994) 332,
hep-th/9308129.}\ref\vafa{
N. Berkovits and C. Vafa,
``On the Uniqueness of String Theory'',
Mod. Phys. Lett. A9 (1994) 653, hep-th/9310170.},
the $N=1$ description of the
RNS superstring can be embedded into a critical $N=2$ string
by defining the twisted $\hat c=2$ $N=2$ superconformal generators as:
\eqn\embed{ T= T_{RNS}, }
$$G^+= \gamma \psi^m \p x_m +
c( \half \p x^m \p x_m +\half \psi^m \p \psi_m  -{3\over 2}
\beta\p\gamma-{\half\gamma\p\beta}-b\p c )-\gamma^2 b +\partial^2 c
+\p (c\xi\eta),$$
$$G^- = b,$$
$$J= cb +\eta\xi,$$
where $T_{RNS}=T_m + T_g$ 
is the $c=0$ stress-tensor of the original RNS matter and
ghost fields, and the $[\beta,\gamma]$
super-reparameterization ghosts have been bosonized \ref\fms
{D. Friedan, E. Martinec, and S. Shenker, ``Conformal Invariance,
Supersymmetry, and String Theory,'' Nucl. Phys. {\bf B271} (1986) 93.}
as
$[\beta= e^{-\phi}\p\xi, \gamma= \eta e^\phi]$.
Note that the zero mode of $G^+$ is the $N=1$ BRST operator and
$J$ is related to the usual ghost current
$j_{ghost}=cb -\gamma\beta$ by 
$J=  j_{ghost} - j_{picture}$ where $j_{picture}$ 
is the picture-current  
defined as $j_{picture}=-\p\phi +\xi\eta $. 

It was shown in \vafa\ that $N=2$ scattering amplitudes computed 
using $N=2$ vertex operators reproduce the usual $N=1$ scattering amplitudes
computed using RNS vertex operators.
The advantage of treating the RNS superstring as an $N=2$ string theory
is that the $N=2$ generators, unlike the original $N=1$ generators,
have no square-root cuts with the spacetime-supersymmetry generators.

In the $-\half$ picture, the spacetime-supersymmetry generator of the
RNS superstring is \fms
\eqn\picture{q_-^A= \oint e^{-\half\phi} S^A}
where $S^A= e^{{i\over 2}(\pm \tau^1 \pm ... \pm\tau^5)}$ 
is the sixteen-component Weyl spin-field constructed by 
bosonizing $\psi^m$ as
$\psi^{2a-2} \pm i\psi^{2a-1} =e^{\pm i\tau^a}$ for $a=1$ to 5.
The index $A$ will sometimes be denoted using U(5) notation 
as $\pm\pm\pm\pm\pm$ where the five choices of $\pm$ correspond to the
five choices of $\pm$ in the definition of $S_A$.
$A$ describes a Weyl spinor when there are an
odd number of $+$'s and an anti-Weyl spinor when there
are an even number of $+$'s. 
In the above bosonization formula, we
have Wick-rotated the spacetime-signature to Euclidean space
so that $\tau^a$ are real chiral bosons satisfying the OPE
$\tau^a(y)\tau^b(z) \to - log(y-z)$.

The algebra generated by the $q^A_-$'s  is 
\eqn\susyminus{
\{q_-^A,q_-^B\}=\sigma^{AB}_m \oint e^{-\phi} \psi^m}
using the OPE's of the chiral bosons $\phi$ and $\tau^a$ where 
$\sigma^{AB}_m$ is the symmetric $16\times 16$ Pauli matrix in ten
dimensions.
\susyminus\ is related to the standard
supersymmetry algebra $\{q^A,q^B\}=\sigma_m^{AB} P^m$ by picture-changing
since $\oint \{Q,\xi\} e^{-\phi} \psi^m =\oint \p x^m$ is the
momentum operator $P^m$ where $\{Q,\xi\} = e^\phi \psi^m \p x_m + ...$
is the picture-changing operator. 
But if one wants the supersymmetry
algebra to close off-shell, one cannot use picture-changing since
the off-shell states are not independent of the location of the
picture-changing operator. 

One can also define the spacetime-supersymmetry generators in the
$+\half$ picture as 
\eqn\pictureplus{q^A_+=\{Q,\xi\} q^A_- =
 \oint (e^{{3\over 2}\phi} b \eta S^A 
+ e^{\half\phi} \p x^m \sigma_m^{AB}\bar S^{\bar B} )}
where $\bar S^{\bar B}$ is an anti-Weyl spin-field.
Although $\{q^A_+,q^B_-\}=\sigma_m^{AB}\oint \p x_m$ without 
picture-changing, this has twice as many supersymmetry generators
as desired, so it is not the $N=1$ $D=10$ supersymmetry algebra.
(It is also not the $N=2$ $D=10$ supersymmetry algebra since
$\{q^A_-,q^B_-\}$ does not vanish.)  
So one needs to find a subset of the
$q_-^A$`s and $q_+^B$'s which generates at least a subset of the
$N=1$ $D=10$ supersymmetry algebra without using
picture-changing.

In previous papers, it was shown how to do this and preserve
manifest $SO(3,1)$ super-Poincar\'e invariance \four\
or manifest
$SO(5,1)$ super-Poincar\'e invariance
\Topo. It will now be shown how to choose
a subset of the supersymmetry generators which preserves U(5)
invariance (after Wick rotation).

\newsec{U(5)-Invariant Description of the Superstring}

Consider the five spacetime-supersymmetry generators defined by
\eqn\five{q^a= \oint e^{-\half\phi} S^a}
where $S^a$ contains one $+$ sign and four $-$ signs in its exponential.
These five generators 
rotate into each other under the $SU(5)\times U(1)$
subgroup of $SO(10)$ rotations which rotate $x^{2a-2} +i x^{2a-1}$
into each other. (There is a Minkowski-space version of this
subgroup, but it is non-compact and has no standard name.)
For later convenience, we shall define 
\eqn\defy{X^a=x^{2a-2} +i x^{2a-1},\quad
\bar X^{\bar a} =x^{2a-2} -i x^{2a-1},}
which satisfy the OPE $X^a(y) \bar X^{\bar b}(z)$
$\to -2\delta^{a\bar b}\log|y-z|$
and which transform respectively as a $(5,1)$ and $(\bar 5,-1)$ under the
$SU(5)\times U(1)$ subgroup. 

Note that $\{q^a,q^b\}=0$,
so it trivially generates the correct supersymmetry algebra without having to
resort to picture-changing. One can still
introduce one more spacetime-supersymmetry
generator without spoiling this property. This generator will be defined
in the $+\half$ picture as
\eqn\one{q^+ = 
 \oint (e^{{3\over 2}\phi} b \eta S^{+++++} 
+{1\over 2} e^{\half\phi}  \bar S^{\bar a} \p X^a )}
where $\bar S^{\bar a}$ is defined to contain four $+$ signs and one $-$ signs
in its exponential.
It is easy to check that $\{q^+,q^a\}={1\over 2}
\oint \p X^a$
and $\{q^+,q^+\}=0$, so one reproduces a subset of the desired supersymmetry 
algebra without
having to use picture-changing.
So six of the spacetime-supersymmetries can be chosen to close off-shell, and
they transform covariantly as a $(5,-3/2)$ and
$(1, 5/2)$ under the $SU(5)\times U(1)$ subgroup of Lorentz transformations.

To make these supersymmetries manifest, one needs to find a field-redefinition
from the RNS worldsheet variables to Green-Schwarz-like
variables which transform simply
under the spacetime-supersymmetry transformations. First, one should
define $\theta^{\bar a}$ and $\theta^-$ variables which satisfy 
$\{q^a,\theta^{\bar b}\}=\delta^{a\bar b}$ and 
$\{q^+,\theta^-\}=1$. This can be
done by defining
$$\theta^{\bar a}
= e^{\half\phi} \bar S^{\bar a},\quad \theta^- =c\xi e^{-{3\over 2}\phi} 
\bar S^{-----}.$$
One then defines the conjugate momentum to these $\t$ variables as
$$p^a= e^{-\half\phi} S^a,\quad p_+ =b\eta e^{{3\over 2}\phi} S^{+++++}.$$
Note that $\t^{\bar a}$ and $\t^+$ have zero conformal weight, transform
under $SU(5)\times U(1)$ as $(\bar 5,3/2)$ and $(1,-5/2)$, and satisfy
the free-field OPE's
$$\ta (y) p^b(z) \to (y-z)^{-1} \delta^{\bar a b},\quad
\t^-(y) p^+(z) \to (y-z)^{-1}.$$
In terms of these variables, the spacetime-supersymmetry generators
of \five\ and \one\ take the simple form
\eqn\simplesusy{q^a =\oint p^a,\quad q^+ =\oint (p^+ +{1\over 2}\ta \p X^a).}

The original RNS variables consisted of $[x^m,\psi^m,b,c,\beta,\gamma]$
and, besides $x^m$, we have defined twelve fermionic variables. 
So there remain two chiral bosons, $\rho$ and $\sigma$,
which need to be defined. Requiring
that they have no singularity with the $\t$'s or $p$'s implies that
they are given by
$$\p\sigma = i(bc + \xi\eta),\quad \p\rho =bc -\xi\eta +3\p\phi
+\sum_{a=1}^5 \psi^{2a-2} \psi^{2a-1}.$$
These chiral bosons satisfy the free-field OPE's
\eqn\opech{\sigma(y)\sigma(z) \to - 2 \log(y-z),\quad
\rho(y)\rho(z) \to -2 \log(y-z)}
and the conformal weight of $e^{im\s+n\rho}$ is $m^2 -m - n^2 -n$.
Furthermore, $e^{im\s +n\rho}$ transforms under $SU(5)\times U(1)$
as a $(1, 5n)$ representation.
The fact that $\rho$ appears without a factor of $i$ in exponentials 
implies that it is a `time-like' chiral boson similar to $\phi$
of the RNS formalism. GSO-projected RNS variables get mapped under
this field-redefinition to exponentials with both $m$ and $n$ integer
or both $m$ and $n$ semi-integer.

Since the new variables obey free-field OPE's, the worldsheet action is
still quadratic and is 
\eqn\action{S= {1\over{2\pi}}
\int d^2 \sigma[\half \p X^a \bar\p \bar X^{\bar a}
+ p^a \bar\p \ta + p^+ \bar\p \t^- 
+ \bar p^a \p \bar\t^{\bar a} + \bar p^+ \p \bar\t^- ]}
where $[\bar\t^{\bar a},\bar \t^-,\bar p^a,\bar p^+]$ are right-moving
variables defined in the same way as their left-moving counterparts.
We have not tried to write the worldsheet action for 
$[\rho,\sigma]$ and $[\bar\rho,\bar\sigma]$ because 
of the usual problems with actions for chiral bosons.
It is straightforward to write the twisted
$N=2$ superconformal generators of \embed\ in terms
of these new variables, and one finds 
\eqn\nequaltwo{T =  \half\p X^a \p \bar X^{\bar a}
 +p^a \p\ta + p^+ \p\t^- +{1\over 4}
 (\p\s\p\s +\p\rho\p\rho)+ \half \p^2 (\rho-i\sigma),}
$$G^+ = e^{\half(\rho +i\s) } [ d^a (\p \bar X^{\bar a}
 +\t^- \p \ta)   -\p\rho \p \t^- -{3\over 2} \p^2\t^-]
+e^{\half(3\rho+i\s)} (d)^5,$$
$$G^- = e^{-\half(\rho+i\s)} p^+,$$
$$J= i\p\s,$$
where
$(d)^5 = {1\over {120}} \e_{abcd} d^a d^b d^c d^d d^e$ and 
\eqn\ddef{d^a = p^a -\half \t^- \p X^a. }
Note that $d^a$ and $\p \bar X^{\bar a}+\t^-\p\ta$ commute
with the spacetime supersymmetry generators $q^a$ and $q^+$,
and since $T$ can be rewritten as 
\eqn\Tsusy{T =  \half\p X^a (\p \bar X^{\bar a} +\t^-\p\ta) 
 +d^a \p\ta + p^+ \p\t^- +{1\over 4}
 (\p\s\p\s +\p\rho\p\rho)+ \half \p^2 (\rho-i\sigma),}
the $N=2$ constraints are manifestly spacetime-supersymmetric.

The U(4)-invariant light-cone Green-Schwarz description is recovered
from this N=2 superconformal field theory
by using the gauge invariances associated with $T$ and $J$ to set 
$\p \bar X^{\bar 1}= 
\rho+i\sigma =0$, and the gauge invariances associated with $G^+$ and $G^-$
to set $\t^{\bar 1} =\t^-=0$. Imposing the N=2 constraints in this gauge
fixes $X^1$,
$p^1$, $p^+$ and $\sigma$, leaving the light-cone degrees of freedom
$[X^j, \bar X^{\bar j}, \t^{\bar j}, p^j]$ where $j=2$ to 5.

\newsec{Physical State Conditions}

As described in \Topo, 
physical states of the superstring can be described by U(1)-neutral
vertex operators, $\Phi$, which satisfy
\eqn\phy{G^+_0 \tilde G^+_0 \Phi = 0,\quad \Phi\sim\Phi  + G^+_0\Omega+
\tilde G^+_0\tilde\Omega}
where $G^+_0$ is the zero mode of the $G^+$ generator (which is the BRST
operator in RNS variables) and
$\tilde G^+_0$ is the zero mode of the generator
$\tilde G^+=[e^{\int J}, G^-].$
In terms of the RNS variables, $\tilde G^+ =\eta$, and in terms of the
Green-Schwarz-like variables, 
\eqn\gtilde{\tilde G^+ = e^{-\half(\rho-i\sigma)} p^+.}
These U(1)-neutral
vertex operators $\Phi$
can be related to the usual physical RNS vertex operators
$V_{RNS}$ by 
\eqn\vrns{V_{RNS}= \tilde G^+_0 \Phi = \eta_0 \Phi;\quad  \Phi=\xi_0 V_{RNS}.}
Note that \phy\ and \vrns\ imply
that $V_{RNS}$ is annihilated by the BRST operator and is in the ``small''
Hilbert space, i.e. it is independent of the $\xi$ zero mode. 

In the RNS formalism, each physical state is represented infinitely
many times in the BRST cohomology because of picture-changing. This
infinite degeneracy is usually removed by fixing all bosons to be in the
zero picture and all fermions to be in the $-\half$ picture. But such a
choice is not manifestly
spacetime-supersymmetric since the generators of \simplesusy\
carry both $\half$ and $-\half$ picture.

An alternative solution
for removing the infinite degeneracy is to require that
all states (both bosons and fermions) carry either $-1$ or $0$ $\rho$-charge,
i.e. their vertex operator is proportional to $e^{n\rho}$ where $n=-1$ or 0.
This solution is manifestly spacetime-supersymmetric since the generators
of \simplesusy\ carry zero $\rho$ charge. It will now be shown that
this solution assigns a unique representative to each physical state.
(A similar solution was used in \ref\field
{N. Berkovits, ``Super-Poincar\'e Invariant Superstring Field
Theory'', Nucl. Phys. B450 (1995) 90, hep-th/9503099.}
for the superstring with manifest
SO(3,1) super-Poincar\'e invariance.)

Physical vertex operators $V_{RNS}$ of the superstring 
must be independent of the $\xi$ zero 
mode, i.e. they must be annihilated by $\oint \eta$ as well as by $Q$.
As mentioned before, there are infinitely many such vertex operators for
each physical state since one can use the picture-raising operator,
$Z=\{Q,\xi\}$, or picture-lowering operator, $Y=c\p\xi e^{-2\phi}$,
to move from one such vertex operator to another one.
However, for each physical state, there is a unique 
vertex operator $\hat V$
which is annihilated by $1-Z$ where $Z$ is the picture-raising operator.  
If $V$ is any physical vertex operator representing this state, then $\hat V$
is given (up to an overall multiplicative factor) by
\eqn\hatv{\hat V = V +\sum_{n=0}^{\infty} Y^n V + 
 \sum_{n=0}^{\infty} Z^n V .}

One can similarly define a unique U(1)-neutral vertex operator $\hat\Phi$ 
for any physical state.
This vertex operator $\hat \Phi$
is defined by 
\eqn\hatphi{ (G^+_0 - \tilde G^+_0)\hat\Phi =0,\quad
\hat\Phi \sim 
\hat\Phi 
+ (G^+_0 - \tilde G^+_0)\hat\Omega,}
and 
can be obtained by
hitting $\hat V$ of $\hatv$ with $\xi_0$.
In terms of the Green-Schwarz-like variables, $\hat\Phi$
can be expanded as $\hat\Phi = \sum_{n=-\infty}^\infty \hat\Phi_n$ where
$\hat\Phi_n$ carries $\rho$-charge $n$. 
It will now be shown that, when $\hat\Phi$ satisfies \hatphi, all components
of $\hat\Phi$ can be determined in terms of $\hat\Phi_0$ and $\hat\Phi_{-1}$.

Using the formulas of \nequaltwo\ and \gtilde, 
$$G^+ - \tilde G^+ = 
- e^{\half(i\sigma -\rho)} p^+ +
 e^{\half(i\sigma +\rho)} 
[d^a (\p \bar X^{\bar a}
 +\t^- \p \ta) 
  -\p\rho \p \t^- -{3\over 2} \p^2\t^-]
+ e^{\half(i\sigma +3\rho)} (d)^5,$$
so one can write 
\eqn\expan{
G^+_0 -\tilde G^+_0 = a_{-\half} + a_{\half} + a_{3\over 2}}
where $a_{-\half}$ is the zero mode of 
$- e^{\half(i\sigma -\rho)} p^+$, $a_\half$ is the zero mode of 
$ e^{\half(i\sigma +\rho)} 
[d^a (\p \bar X^{\bar a}
 +\t^- \p \ta) $
$  -\p\rho \p \t^- -{3\over 2} \p^2\t^-],$
and $a_{3\over 2}$ is the zero mode of
$ e^{\half(i\sigma +3\rho)} 
(d)^5.$ It will be useful to
note that the cohomologies of 
$a_{-\half}$  and 
$a_{3\over 2}$ are trivial. This is easy to
show since $
a_{-\half} F =0$ implies that 
$F= 
a_{-\half} ( 
- e^{-\half(i\sigma-\rho)} \t^- F)$ and 
$a_{3\over 2} F =0$ implies that 
$F= 
a_{3\over 2} ( 
e^{-\half(i\sigma+3\rho)} (\t)^5 F)$
where $(\t)^5 = {1\over{120}}\e^{abcde}\ta\t^{\bar b}
\t^{\bar c}\t^{\bar d}\t^{\bar e}$.

Using $(G^+_0 -\tilde G^+_0)\hat\Phi=0$, one can show that 
\eqn\eqmo{a_{3\over 2} \hat\Phi_{-2} = - a_{\half} \hat\Phi_{-1} - 
a_{-\half} \hat\Phi_0  ,\quad
a_{-\half} \hat\Phi_{1} = - a_{\half} \hat\Phi_{0} - 
a_{3\over 2} \hat\Phi_{-1}.}
These equations are invariant under the gauge transformations coming
from
$\delta\hat\Phi = 
(G^+_0 -\tilde G^+_0)\hat\Omega$, 
\eqn\gthat{\delta \hat\Phi_{-2} = a_{3\over 2} \hat\Omega_{-{7\over 2}},\quad
\delta \hat\Phi_{-3} = a_{1\over 2} \hat\Omega_{-{7\over 2}},\quad
\delta \hat\Phi_{-4} = a_{-\half} \hat\Omega_{-{7\over 2}},}
$$\delta \hat\Phi_{1} = a_{-\half} \hat\Omega_{{3\over 2}},\quad
\delta \hat\Phi_{2} = a_{1\over 2} \hat\Omega_{{3\over 2}},\quad
\delta \hat\Phi_{3} = a_{3\over 2} \hat\Omega_{{3\over 2}}$$
where $\hat\O=\sum_{n+\half=-\infty}^\infty \hat\O_n.$
Since the cohomologies of $a_{3\over 2}$ and $a_{-\half}$ are trivial, 
one can use \eqmo\ to
express $\hat\Phi_{-2}$ and $\hat\Phi_1$ in terms
of $\hat\Phi_{-1}$ and $\hat\Phi_{0}$. Similarly, one can show that
$\hat\Phi_n$ for all $n$
can be expressed in terms of $\hat\Phi_{-1}$ and $\hat\Phi_{0}$.

The equations of \eqmo\ imply that $\hat\Phi_0$ and $\hat\Phi_{-1}$ satisfy
\eqn\eqm{a_{3\over 2}( a_{\half} \hat\Phi_{-1} +
a_{-\half} \hat\Phi_0 )=0 ,\quad
a_{-\half} ( a_{\half} \hat\Phi_{0} + 
a_{3\over 2} \hat\Phi_{-1}) =0.}
These equations are invariant under the gauge transformations 
\eqn\gt{\delta \hat\Phi_{-1} = a_{3\over 2} \hat \O_{-{5\over 2}} + 
a_{\half}\hat\O_{-{3\over 2}} + a_{-\half} \hat\O_{-\half},\quad 
\delta \hat\Phi_{0} = a_{3\over 2} \hat \O_{-{3\over 2}} + 
a_{\half}\hat\O_{-{1\over 2}} + a_{-\half} \hat\O_{\half}.}
So any physical state of the superstring is uniquely described
by the fields $\hat\Phi_0$ and $\hat\Phi_{-1}$, which satisfy the equations
of 
\eqm, and which are defined up to the gauge transformations of \gt.
It should be noted that $\Phi=\hat\Phi_{-1}+\hat\Phi_0$ does not
satisfy \phy, and therefore is not an acceptable physical vertex operator
for computing scattering amplitudes. However, one can construct
a physical vertex operator from $\hat\Phi_{-1}$ and $\hat\Phi_0$ 
as
\eqn\pvo{\Phi = a_{\half}[(\t)^5 e^{-\half(i\sigma+3\rho)}\hat\Phi_{-1}]
+\hat\Phi_{-1} + \hat\Phi_0 - a_{3\over 2} [\t^- e^{-\half(i\sigma-\rho)}\hat
\Phi_{-1}],}
which, using \eqm\ and \gt, can be shown to satisfy \phy.

It will now be shown that \eqm\ and \gt\ correctly describe 
the physical degrees
of freedom for the massless states of the open superstring.
For massless states, the vertex operator must have conformal weight zero
and zero momentum. Since it is U(1)-neutral, the most general
such vertex operator with $\rho$-charge $-1$ or $0$ is 
$$\hat\Phi_{-1} = e^{-\rho} B( x^m,\ta,\t^-), \quad
\hat\Phi_{0} = C( x^m,\ta,\t^-), $$
where $B$ and $C$ are two superfields depending on the ten $x$'s and six
$\t$'s. 

The equations of \eqm\ imply that
\eqn\eqmless{\N_{\bar b}\N_{\bar c}\N_{\bar d} 
(2 \N_{\bar a}\p_a B+ \N_- C)=0, \quad
-\half (\N^4)_a \p_{\bar a} B + (\N)^5 \N_- B + \N_{\bar a}\N_- \p_a C=0}
where $\N_- = \p/\p\t^-$, $\N_{\bar a}=\p/\p\ta +\t^-\p_{\bar a}$,
$\p_a = \p/\p X^a,$
$\p_{\bar a}=\p/\p \bar X^{\bar a}$, 
$(\N^4)_a 
={1\over{24}}\e_{abcde}\N_{\bar b}\N_{\bar c}\N_{\bar d}\N_{\bar e},$
and 
$(\N)^5 
={1\over{120}}\e_{abcde}\N_{\bar a }
\N_{\bar b}\N_{\bar c}\N_{\bar d}\N_{\bar e}.$
These equations are invariant under the gauge transformations
\eqn\gtless{\delta B= \N_{\bar a}\N_{\bar b} \omega^{\bar a\bar b}
-2 \N_{\bar a}\p_a \lambda,\quad 
\delta C= (\N)^5 \lambda - \N_-\Sigma}
which come from choosing the gauge parameters of \gt\ to be
$$\hat\Omega_{-{5\over 2}}=
e^{-\half(i\sigma+5\rho)}
\e^{abcde} \p\ta\p\tb\p\t^{\bar c} \omega^{\bar d
\bar e},\quad \hat\Omega_{-{3\over 2}}= e^{-\half(i\sigma+3\rho)}\lambda,\quad
\hat\Omega_{{1\over 2}}= e^{-\half(i\sigma-\rho)}\Sigma.$$
Note that $\hat\Omega_{-\half}$ has no contribution at the massless level.

Using the gauge transformations of \gtless, one can algebraically
gauge $B$ and $C$ into
the form
\eqn\gab{ B= -2(\t^4)^a A_a + (\t)^5 \Psi + \t^- (\t^4)^a \Psi_a
- \t^- (\t^5) D,}
$$C= \t^- [ \ta A_{\bar a} +\ta\tb\Psi_{\bar a\bar b} + 
\ta\tb\t^{\bar c} N_{\bar a\bar b\bar c}
+ (\t^4)^a \xi_a + (\t)^5 P]$$
where $(\t^4)^a = {1\over 24}\e^{abcde} 
\tb\t^{\bar c}\t^{\bar d}\t^{\bar e}$   
and all fields on the right-hand side of \gab\ are component fields 
depending only on $x^m$. There is still one remaining gauge parameter
given by $\lambda = (\t)^5 c$ and $\Sigma= \t^- c$, which transforms
the component fields of \gab\ as 
$$\delta A_a = \p_a c, \quad \delta A_{\bar a} =\p_{\bar a} c, \quad
\delta D = 2 \p_a\p_{\bar a} c.$$

It is straightforward to check that the equations of \eqmless\ imply
that 
\eqn\comp{D = \p_a A_{\bar a} + \p_{\bar a} A_a,\quad
\p_a D = 2 \p_b \p_{\bar b} A_a,\quad
\p_{\bar a} D = 2 \p_b \p_{\bar b} A_{\bar a},}
$$4 \e_{\bar a\bar b\bar c\bar d\bar e} 
\p_a \Psi_b = -\p_{[\bar c}\Psi_{\bar d
\bar e]},\quad \p_{\bar a}\Psi = - 4\p_b \Psi_{\bar a\bar b},\quad
\p_{\bar a}\Psi_a =0,$$
$$N_{\bar c\bar d\bar e}=-{2\over 3}\e_{\bar a\bar b\bar c\bar d\bar e}
\p_a A_b, \quad \xi_a = -2 \p_a \Psi,\quad P=0.$$
The first line of \comp\ implies that $A_a$ and $A_{\bar a}$ can be
combined into a ten-component
vector $A_m$ satisfying Maxwell's equation $\p^m \p_{[m} A_{n]} =0$.
The second line of \comp\ implies that $\Psi$, $\Psi_a$ and
$\Psi_{\bar d\bar e}$ can be combined into a sixteen-component anti-Weyl
spinor
$\Psi^{\bar A}$ satisfying the Dirac equation $\p^m \s_m^{AB} \Psi^{\bar B}=0.$
So the physical content of the massless states of the open superstring
has been correctly described. 

\vskip 10pt
{\bf Acknowledgements:} The author would like to thank CNPq grant
300256/94-9 for partial financial support.

\listrefs

\end